\documentclass[preprint,showpacs,showkeys,preprintnumbers,amsmath,amssymb]{revtex4}
\usepackage{graphicx}
\usepackage{dcolumn}
\usepackage{bm}
\begin{document}
\preprint{APS/123-QED}
\title{Phase diagram of the spin-1 XXZ Heisenberg ferromagnet with a single-ion anisotropy}
\author{Jozef Stre\v{c}ka} 
\email{jozef.strecka@upjs.sk} 
\homepage{http://158.197.33.91/~strecka}
\author{J\'an Dely}
\author{Lucia \v{C}anov\'a}
\affiliation{Department of Theoretical Physics and Astrophysics, 
Faculty of Science, \\ P. J. \v{S}af\'{a}rik University, Park Angelinum 9,
040 01 Ko\v{s}ice, Slovak Republic}

\date{\today}
             
\begin{abstract}
Phase diagram of the spin-1 quantum Heisenberg model with both exchange as well as single-ion anisotropy is constructed within the framework of pair approximation formulated as a variational procedure based on the Gibbs-Bogoliubov inequality. In this form adapted variational approach is 
used to obtain the results equivalent with the Oguchi's pair approximation. It is shown that the single-ion anisotropy induces a tricritical behaviour in the considered model system and a location 
of tricritical points is found in dependence on the exchange anisotropy strength.  
\end{abstract}

\maketitle

\section{Introduction}
\label{intro}

Phase transitions and critical phenomena of quantum spin systems currently attract a great deal of 
interest \cite{sach99}. As usual, the quantum Heisenberg model is used as a basic generating model 
which should be appropriate for investigating quantum properties of insulating magnetic materials
\cite{gat85,car86,kah93,gat99}. However, a rigorous proof known as the Mermin-Wagner theorem \cite{mer66} prohibits a spontaneous long-range order for the isotropic spin-1/2 Heisenberg model on the one- and two-dimensional lattices and hence, the spontaneous ordering might in principle appear either if a three-dimensional magnetic structure is considered \cite{dys76} or a non-zero magnetic anisotropy is involved in the studied model Hamiltonian \cite{fro77}. On the other hand, it is currently well established that obvious quantum manifestations usually arise from a mutual combination of several factors, especially, when the low-dimensional magnetic structure is combined with as low coordination number as possible and low quantum spin number. Apparently, these opposite trends make hard to find a long-range ordered system that simultaneously exhibits evident quantum effects. Investigation of quantum spin systems, which can exhibit a non-trivial criticality, thus remains among the most challenging tasks in the statistical and solid-state physics.  

Over the last few decades, there has been increasing interest in the study of the effect of different
anisotropies (single-ion, Dzyaloshinskii-Moriya, exchange) on the critical behaviour of the spin-1  
quantum Heisenberg ferromagnet. The main interest to study this model system arises since Stanley 
and Kaplan \cite{sta66,sta67} proved the existence of a phase transition in the two- and three-dimensional Heisenberg ferromagnets. In addition, the ferromagnetic quantum Heisenberg model 
with the spin-1 has relevant connection with several nickel-based coordination compounds, 
which provide excellent experimental realization of this model system \cite{def,djm}. Up to now, 
the spin-1 quantum Heisenberg model has been explored within the standard mean-field approximation \cite{tag74,buz88}, random phase approximation \cite{mic77} or linked-cluster expansion \cite{pan93,pan95}. By making use of the pair approximation \cite{ury80,iwa97,lu06,sun06}, 
several further studies have been concerned with the critical behaviour of the anisotropic spin-1 
XXZ Heisenberg ferromagnet with bilinear and biquadratic interactions \cite{ury80,iwa97}, the isotropic spin-1 Heisenberg ferromagnet with the bilinear and biquadratic interactions and the single-ion anisotropy \cite{lu06}, as well as, the anisotropic spin-1 XXZ Heisenberg ferromagnet with an antisymmetric Dzyaloshinskii-Moriya interaction \cite{sun06}. 
To the best of our knowledge, the critical properties of the spin-1 XXZ Heisenberg ferromagnet
with the uniaxial single-ion anisotropy have not been dealt with in the literature yet. Therefore, 
the primary goal of present work is to examine this model system which represents another eligible candidate for displaying an interesting criticality affected by quantum fluctuations.  

The rest of this paper is organized as follows. In Section \ref{model}, we briefly describe 
the model system and basic steps of the variational procedure which gives results equivalent to the Oguchi's pair approximation \cite{ogu55}. Section \ref{result} deals with the most interesting numerical results obtained for the ground-state and finite-temperature phase diagrams. Magnetization dependences on the temperature, for several values of exchange and single-ion anisotropies, are also displayed in Section \ref{result}. Finally, some concluding remarks are drawn in Section \ref{conclusion}.  

\section{Model and method}
\label{model}
Let us consider the Hamiltonian of the spin-1 quantum Heisenberg model: 
\begin{eqnarray}
{\cal H} = - J \sum_{(i,j)}^{Nq/2} [\Delta (S_i^x S_j^x + S_i^y S_j^y) + S_i^z S_j^z] 
           - D \sum_{i=1}^{N} (S_i^z)^2 - H \sum_{i=1}^{N} S_i^z, 
\label{ham}
\end{eqnarray}
where $S_i^{\alpha}$ ($\alpha=x,y,z$) denotes spatial components of the spin-1 operator at the lattice site $i$, the first summation runs over nearest-neighbour pairs on a lattice with a coordination 
number $q$ and the other two summations are carried out over all $N$ lattice sites. The first 
term in Hamiltonian (\ref{ham}) labels the ferromagnetic XXZ Heisenberg exchange interaction with the coupling constant $J>0$, $\Delta$ is the exchange anisotropy in this interaction, the parameter $D$ stands for the uniaxial single-ion anisotropy and the last term incorporates the effect of external magnetic field $H$. 

The model system described by means of the Hamiltonian (\ref{ham}) will be treated within 
the pair approximation formulated as a variational procedure based on the Gibbs-Bogoliubov 
inequality \cite{bog47,fey55,bog62,fal70}:
\begin{eqnarray}
G \leq G_0 + \langle {\cal H} - {\cal H}_0 \rangle_0.
\label{gbf}
\end{eqnarray}
Above, $G$ is the Gibbs free energy of the system described by the Hamiltonian (\ref{ham}), $G_0$ 
is the Gibbs free energy of a simplified model system given by a trial Hamiltonian ${\cal H}_0$, 
and $\langle \ldots \rangle_0$ indicates a canonical ensemble averaging performed within this 
simplified model system. Notice that the choice of the trial Hamiltonian ${\cal H}_0$
is arbitrary, however, its form directly determines an accuracy of the obtained results. 
If only single-site interaction terms are included in the trial Hamiltonian, i.e. single-spin 
cluster terms are used as the trial Hamiltonian, then, one obtains results equivalent to the 
mean-field approximation. Similarly, if a two-spin cluster Hamiltonian is chosen as the trial Hamiltonian, the obtained results will be equivalent to the Oguchi's pair approximation, which 
is superior to the mean-field approach. 

In the present work, we shall employ the two-spin cluster approach for the considered model system
in order to obtain results equivalent to the Oguchi's pair approximation \cite{ogu55}. 
The two-spin cluster trial Hamiltonian can be written in this compact form:
\begin{eqnarray}
{\cal H}_0 &=& \sum_{k=1}^{N/2} {\cal H}_k, \label{trial1} \\
{\cal H}_k &=& - \lambda [\delta (S_{k1}^x S_{k2}^x + S_{k1}^y S_{k2}^y) + S_{k1}^z S_{k2}^z] 
        \nonumber\\ &&- \eta [(S_{k1}^z)^2 + (S_{k2}^z)^2] - \gamma (S_{k1}^z + S_{k2}^z), 
\label{trial2}
\end{eqnarray}
where the first summation is carried out over $N/2$ spin pairs and $\lambda$, $\delta$, $\eta$, and 
$\gamma$ denote variational parameters which have obvious physical meaning. It is noteworthy 
that an explicit expression of the variational parameters can be obtained by minimizing the 
right-hand-side of Eq. (\ref{gbf}), i.e. by obtaining the best estimate of the true Gibbs free 
energy. Following the standard procedure one easily derives:
\begin{eqnarray}
\lambda = J, \quad \delta = \Delta, \quad \eta = D, \quad \gamma = (q - 1) J m_0 + H,
\label{para}
\end{eqnarray}
where $m_0 \equiv \langle S_i^z \rangle_0$ denotes the magnetization per one site of the set of independent spin-1 dimers described by means of the Hamiltonian ${\cal H}_0$. By substituting 
optimized values of the variational parameters (\ref{para}) into the inequality (\ref{gbf}) one consequently yields the best upper estimate of the true Gibbs free energy within the 
pair-approximation method: 
\begin{eqnarray}
G = \frac{N}{2} G_k + \frac{N J}{2} (q-1) m_0^2.
\label{gfe}
\end{eqnarray}
Above, $G_k$ labels the Gibbs free energy of the spin-1 Heisenberg dimer given by the Hamiltonian (\ref{trial2}). With the help of Eq. (\ref{gfe}), one can straightforwardly verify that the magnetization of the original model directly equals to the magnetization of the corresponding 
dimer model, i.e. $m \equiv \langle S_i^z \rangle = \langle S_i^z \rangle_0 \equiv m_0$. 
Of course, similar relations can be established for another quantities, as well.  

To complete solution of the model under investigation, it is further necessary to calculate 
the Gibbs free energy, magnetization and other relevant quantities of the corresponding spin-1 
dimer model given by the Hamiltonian (\ref{trial2}). Fortunately, an explicit form of all relevant quantities (Gibbs free energy, magnetization, correlation functions, quadrupolar moment) can be 
found for this model system elsewhere \cite{str05}. Referring to these results, the solution 
of the considered model system is formally completed. For the sake of brevity, we just merely 
quote final expressions for the Gibbs free energy $G_k$ and the magnetization $m_0$, both entering 
into Eq. (\ref{gfe}):
\begin{eqnarray}
G_k &=& - \beta^{-1} \ln Z_k, \label{fin1} \\
Z_k &=& 2 \exp[\beta (\lambda + 2 \eta)] \cosh(2 \beta \gamma) 
            + 4 \exp(\beta \eta) \cosh(\beta \gamma) \cosh(\beta \lambda \delta) \nonumber\\
	    &+& \exp[\beta (2\eta - \lambda)] + 2 \exp[\beta (\eta- \lambda / 2)] \cosh(\beta W), \label{fin2} \\        
m_0 &=& \frac{1}{Z_d} \{ 2 \exp[\beta(\lambda + 2 \eta)] \sinh(2 \beta \gamma) 
            + 2 \exp(\beta \eta) \sinh(\beta \gamma) \cosh(\beta \lambda \delta) \}, \label{fin3}    
\end{eqnarray}
where $W = \sqrt{(\eta- \lambda/2)^2 + 2 (\lambda \delta)^2}$, $\beta = 1/(k_{\rm B} T)$, $k_{\rm B}$ 
is the Boltzmann's constant, $T$ labels the absolute temperature, and the variational parameters $\lambda$, $\delta$, $\eta$, and $\gamma$ take their optimized values (\ref{para}). It is quite 
evident that the magnetization $m_0$ must obey the self-consistent transcendental Eq. (\ref{fin3}) (recall that it enters into the variational parameter $\gamma$ given by Eq. (\ref{para})), which 
might possibly have more than one solution. Accordingly, the stable solution for the magnetization 
$m_0$ is the one that minimizes the overall Gibbs free energy (\ref{gfe}). 

In an absence of the external magnetic field ($H=0$), the magnetization tends gradually to zero 
in the vicinity of a continuous (second-order) phase transition from the ordered phase ($m=m_0 \neq 0$) 
towards the disordered phase ($m=m_0 = 0$). According to this, the magnetization (\ref{fin3}) 
close to the second-order phase transition can be expanded into the series:  
\begin{eqnarray}
m = a m + b m^3 + c m^5 + \ldots.
\end{eqnarray}
Notice that the coefficients $a$, $b$, and $c$ depend on the temperature and all parameters involved 
in the model Hamiltonian (\ref{ham}). Then, the power expansion of the magnetization $m$ can be straightforwardly used to locate second-order transition lines and tricritical points by following 
the standard procedure described in several previous works \cite{ben85,kan86,tuc89,cha92,jia93}. 
The critical temperatures corresponding to the second-order transitions must obey the condition 
$a = 1$, $b < 0$, while the tricritical points can be located from the constraint $a = 1$, $b = 0$, 
and $c < 0$. Finally, the critical temperatures of discontinuous (first-order) transitions must 
be obtained from a comparison of Gibbs free energy of the lowest energy ordered phase 
with the Gibbs free energy of the disordered phase.   

\section{Results and discussion}
\label{result}

Before proceeding to a discussion of the most interesting numerical results, let us firstly mention 
that some particular results for the considered model system have already been reported on by the present authors elsewhere \cite{del06}. Note that in the former preliminary report we have used an alternate approach based on the original Oguchi's pair approximation to study a particular case with the coordination number $q=4$ corresponding  to the square and diamond lattices. In the present article, we shall further focus our attention to other particular case with the coordination number $q=6$, which corresponds to the case of the triangular and simple-cubic lattices. A brief comparison with the results obtained previously will be made in conclusion.

Now, let us take a closer look at the ground-state behaviour. A detailed analysis of our numerical results shows that the ground-state phase boundary between the ferromagnetically ordered and 
the disordered phases can be allocated with the aid of following condition:
\begin{eqnarray}
\frac{D_{\rm b}}{J} = - \frac{q}{2} + \frac{\Delta^2}{q+1}.
\label{gs}
\end{eqnarray}
It is quite obvious from the Eq. (\ref{gs}) that the ground-state phase boundary between 
the ordered and disordered phases shifts to the more positive (weaker) single-ion anisotropies 
when the parameter $\Delta$ is raised from zero. As a matter of fact, the order-disorder 
transition moves towards the weaker single-ion anisotropies for any $\Delta \neq0 $ in comparison 
with the result $D/J = - q/2$ attained in the semi-classical Ising limit ($\Delta = 0$). 
This result is taken to mean  that a destabilization of the ferromagnetic order originates from 
raising quantum fluctuations, which work in conjunction with the single-ion anisotropy in the 
view of destroying of the ferromagnetic long-range order at zero temperature. It is worthwhile 
to remark that an appearance of the planar (XY) long-range ordering cannot be definitely ruled out 
in the parameter space with predominant easy-plane interactions ($D<0$ and/or $\Delta>1$), where 
we have found the disordered phase only. It should be stressed, however, that the present form of two-spin cluster mean-field treatment cannot resolve a presence of the ferromagnetic long-range 
order inherent to XY-type models \cite{lie62,dys78,ken88,kub88} unlike the conventional Ising-like ferromagnetic long-range order with only one non-zero component of the spontaneous magnetization.
				
Next, let us turn our attention to the finite-temperature phase diagram, which is shown in Fig. 1 
in the reduced units $d = D/J$ and $t = k_{\rm B}T/J$ for the simple-cubic (triangular) lattice and different values of the exchange anisotropy $\Delta$. In this figure, the solid and dashed lines represent second- and first-order phase transitions between the ferromagnetic and paramagnetic phases, respectively, while the black circles denote positions of the tricritical points. It is quite obvious from this figure that the considered model system exhibits the highest values of critical temperature
\begin{figure}[h]
\begin{center}
\includegraphics[width=100mm]{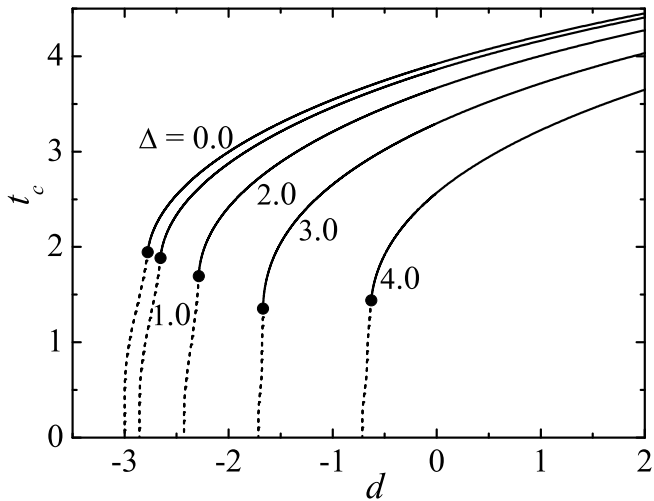}
\end{center}
\vspace{-15mm}
\caption{The phase diagram of the spin-1 Heisenberg model for the simple-cubic (triangular) lattice 
and several values of the exchange anisotropy $\Delta$. The solid and dashed lines represent second- and first-order phase transitions, respectively. The black circles denote positions of the tricritical points.}
\end{figure}
in the Ising limit ($\Delta = 0$). The gradual increase of the exchange anisotropy $\Delta$ reduces 
the transition temperature as a result of raising quantum fluctuations. It is worthwhile to 
remark that all the lines of second-order phase transitions, for arbitrary but finite $\Delta$, 
have the same asymptotic behaviour in the limit $d \to \infty$. Actually, the critical temperature 
of the continuous transitions does not depend on the exchange anisotropy in this limiting case and 
it is equal to $t^* = 5.847$. Moreover, it should be also mentioned that our approach yields for 
the Ising case without the single-ion anisotropy ($\Delta = 0$, $d = 0$) the critical temperature 
$t_c = 3.922$, which is consistent with the result of other pair-approximation methods \cite{sun06} 
and is simultaneously superior to the result $t_c = 4.0$ obtained from the standard mean-field approximation \cite{fit92}. In addition, it can be clearly seen from Fig. 1 that the transition temperature of the continuous phase transition monotonically decreases by decreasing the single-ion anisotropy $d$ until the tricritical point (TCP) is reached. Further decrease of the anisotropy parameter $d$ changes the second-order phase transitions towards the first-order ones. It should 
be realized, nevertheless, that the first-order phase transitions occur merely in a narrow region 
of single-ion anisotropies close to the boundary value (\ref{gs}) at which both completely ordered phases with $m = \pm 1$ have the identical energy (coexist together) with the disordered phase 
with $m=0$ and one asymptotically reaches the first-order phase transition between them in the zero-temperature limit. An origin of discontinuous phase transitions could be therefore related 
to the fact that the ordered and disordered phases have very close energies near the boundary single-ion anisotropy (\ref{gs}) (the former ones are being slightly lower in energy) and the small 
temperature change might possibly induce a phase coexistence (energy equivalence) between them, 
what consequently leads to the discontinuous phase transition. In Fig. 2, we depict more clearly 
the position of TCPs in dependence on the single-ion and exchange anisotropies by the
\begin{figure}
\begin{center}
\includegraphics[width=100mm]{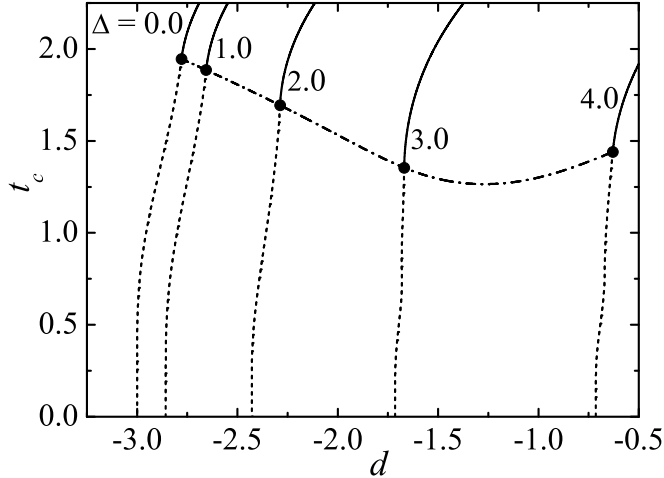}
\end{center}
\vspace{-15mm}
\caption{The phase diagram of the spin-1 Heisenberg model for $q = 6$ and 
$\Delta = 0.0, 1.0, 2.0, 3.0,$ and $4.0$. The solid and dashed lines represent second- 
and first-order phase transitions, respectively. The black circles denote positions of 
the tricritical points. The dot-and-dash line represents the location of
tricritical points in dependence on the exchange anisotropy $\Delta$.}  
\end{figure}
dot-and-dash line in order to clarify how the type of phase transition changes with 
the anisotropy parameters. As one can see from this figure, the $d$-coordinate of TCPs ($d_t$) 
shifts to more positive values upon strengthening of $\Delta$, while the $t$-coordinate of 
TCPs behaves as a non-monotonic function of the exchange anisotropy $\Delta$ with a minimum 
at $\Delta_{min} = 3.459$. 

To illustrate the effect of uniaxial single-ion anisotropy on the phase transitions, the thermal variation of the magnetization $m$ is shown in Fig. 3 for the case of isotropic spin-1 Heisenberg 
model ($\Delta = 1.0$) and several values of $d$. It can be clearly seen from this figure that the reduction of the single-ion anisotropy causes lowering of the critical temperature $t_c$. Furthermore,
\begin{figure}
\begin{center}
\includegraphics[width=100mm]{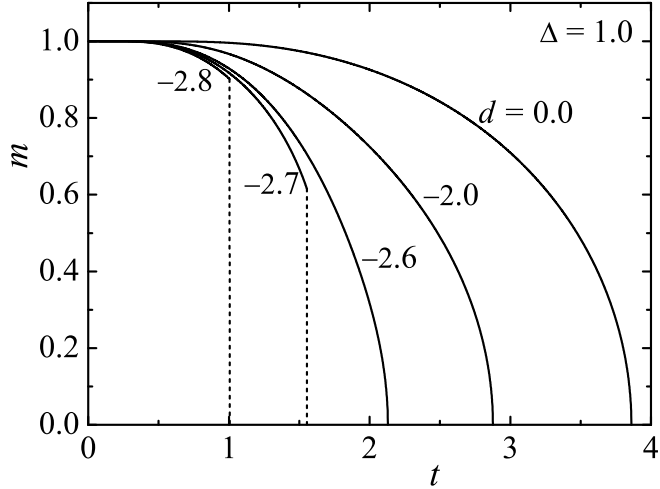}
\end{center}
\vspace{-15mm}
\caption{The temperature dependence of the magnetization $m$ for the isotropic spin-1 Heisenberg model ($\Delta$ = 1.0) on the simple-cubic (triangular) lattice, when the value of the single-ion anisotropy parameter $d$ changes. The dashed lines represent the discontinuities of the magnetization at the first-order phase transitions.}  
\end{figure}
 it is also evident that the magnetization varies smoothly to zero for $d = 0.0$, $-2.0$, and $-2.6$ until the temperature reaches its critical value. This behaviour of magnetization, which is typical 
for the second-order (continuous) phase transitions, persists until $d > d_t$ ($d_t = -2.656$ for $\Delta = 1.0$ and $q=6$). On the other hand, the magnetization jumps discontinuously to zero for $d<d_t$ (e.g. see the curves  for $d=-2.7$ and $-2.8$), what is characteristic feature of the first-order (discontinuous) phase transitions. As one can see, this discontinuity in the magnetization 
increases rather abruptly as the single-ion anisotropy moves to more negative values with respect 
to the $d_t$ value. Finally, it should be pointed out that the similar variations of magnetization curves occur for any value of the exchange anisotropy $\Delta$.           

\section{Conclusion}
\label{conclusion}

In the present paper, the phase diagram of the anisotropic spin-1 XXZ Heisenberg model with the uniaxial
single-ion anisotropy is examined within the variational procedure based on the Gibbs-Bogoliubov inequality,
which gives results equivalent to the Oguchi's pair approximation \cite{ogu55}. A comparison between the results obtained in the present study and those attained within the standard Oguchi approximation actually implies an equivalence between both the methods. The most important benefit of using the variational approach based on the Gibbs-Bogoliubov inequality is that in this way adapted method enables obtaining of all thermodynamic quantities in a self-consistent manner and moreover, it is also well suited to discern the continuous phase transitions from the discontinuous ones by distinguishing of the stable, metastable and unstable solutions inherent to the approximation used. 

In the spirit of the applied pair-approximation method we have demonstrated that the single-ion anisotropy as well as the exchange anisotropy have a significant influence on the critical behaviour and both these anisotropy parameters can cause a tricritical phenomenon, i.e. the change of the continuous phase transition to the discontinuous one. Our results can serve in evidence that the tricritical phenomenon may occur in the investigated model system if at least one of the anisotropy parameters provides a sufficiently strong source of the easy-plane anisotropy. Note furthermore 
that the obtained results are rather general in that they are qualitatively independent of the 
lattice coordination number. The comparison between the results to be presented in this work with 
those reported on previously for other particular case \cite{del06} actually implies that the 
model under investigation shows qualitatively the same features irrespective of the lattice coordination number.   

\begin{center}
\begin{acknowledgments}
This work was supported under the grants Nos. VVGS 11/2006, VEGA 1/2009/05 and APVT 20-005204.
\end{acknowledgments}
\end{center}

\end{document}